\documentclass[amsmath,amssymb,nofootinbib,showpacs,twocolumn,nofootinbib]{revtex4-1} 
\usepackage{dcolumn}
\usepackage{bm}
\usepackage{graphicx}
\usepackage{epsfig}
\usepackage{color}
\usepackage{natbib}
\usepackage{twoopt}
\usepackage{dashrule}

\definecolor{mycolor}{rgb}{0.45, 0.31, 0.59}
\usepackage[breaklinks=true,colorlinks=true,linkcolor=mycolor,citecolor=mycolor,urlcolor=mycolor,pdfauthor={Tomassetti},pdftitle={ProtonMFP}]{hyperref}

\usepackage{xspace}

\newcommand{\GCR}{GCR\xspace}
\newcommand{\GCRs}{GCRs\xspace}

\newcommand{\etal}{et alii}
 
\newcommand{\ie}{\textit{i.e.}}

\newcommand{\ApJ}{Astrophys. J.}

\newcommand{\PRL}{Phys. Rev. Lett.}
\newcommand{\PRD}{Phys. Rev. D}
\newcommand{\ASR}{Adv. Space Res.}

\def\Journal#1#2#3#4{{#1}\,{#2}, #3 (#4)}
\def\href#1#2{#2;}

\def\citep#1{\cite{#1}}

\usepackage{verbatim}

\begin{document}

\title{Data driven analysis of cosmic rays in the heliosphere: diffusion of cosmic protons}

\author{N. Tomassetti$^{\,1}$, E. Fiandrini$^{\,1}$, B. Bertucci$^{\,1}$, F. Donnini$^{\,2}$, M. Graziani$^{\,1}$, B. Khiali$^{\,3}$, A. Reina Conde$^{\,4}$}
\address{$^{1}$\,Dipartimento di Fisica e Geologia, University of Perugia, Italy}
\address{$^{2}$\,INFN - Sezione di Perugia, Italy}
\address{$^{3}$\,INFN - Sezione di Roma Tor Vergata, Italy}
\address{$^{4}$\,INFN - Sezione di Bologna, Italy}

\begin{abstract}
    Understanding the time-dependent relationship between the Sun's variability and cosmic rays (\GCR) is
  essential for developing predictive models of energetic radiation in space.
  When traveling inside the heliosphere, \GCRs are affected by magnetic turbulence and solar wind disturbances which
  result in the so-called solar modulation effect. To investigate this phenomenon, we have performed a data-driven analysis
  of the temporal dependence of the \GCR flux over the solar cycle. With a global statistical inference of \GCR data collected
  in space by AMS-02 and PAMELA on monthly basis, we have determined the rigidity and time dependence of the \GCR diffusion mean free path.
  Here we present our results for \GCR protons, we discuss their interpretation in terms of basic processes
  of particle transport and their relations with the dynamics of the heliospheric plasma.
\end{abstract}

\maketitle

\section{Introduction}    
\label{Sec::Introduction} 

When entering the heliosphere, Galactic Cosmic Rays (\GCRs) are subjected to the solar modulation effect which causes a
significant modification in the energy spectrum of their flux in comparison with the local interstellar spectrum (LIS) outside the heliosphere.
To understand solar modulation, it is crucial to model the transport processes of \GCRs in the solar wind and its embedded magnetic field. 
The main processes are diffusion, drift, advection and adiabatic deceleration.
All these processes are time dependent and follow the quasiperiodical 11-year solar cycle. 
Solar modulation is very important in \GCR physics and heliophysics \cite{Moraal2013,Potgieter2013}.  
Modeling the temporal evolution of \GCRs in interplanetary space is also important for assessing radiation risks
and hazards in long-duration crewed space missions.
In this respect, the recent high-precision and time-resolved data from AMS-02 \cite{Aguilar2018PHeVSTime,Aguilar2018LeptonVSTime}
and PAMELA \cite{Adriani2013,Martucci2018} experiments offer a unique possibility to study the \GCR modulation over a long period of time.

\section{Methodology}     
\label{Sec::Methodology}  

\subsection{The model} 
\label{Sec::Model}     

We implemented a 2D description of the heliosphere, modeled as a spherical bubble centered on the Sun. 
The wind flows radially from the Sun, with a speed $V_{sw}(r,\theta,t)$ that depends on 
helioradius $r$, heliolatitude $\theta$, and time $t$ \citep{Fiandrini2021}.
The solar wind drops to a subsonic speed across the termination shock  $r_{\rm{TS}}=85$\,AU,
and vanishes at the heliopause $r_{\rm{HP}}=122$\,AU. The Earth is placed in the equatorial plane, at $r_{0}=$1\,AU.
The interplanetary magnetic field (IMF) $vec{B}$ is wound up in a rotating spiral,
where its angular aperture depends on the wind speed.
Similarly, the heliospheric current sheet (HCS) is modeled on that structure.
The HCS is a rotating layer which divides the IMF into two hemispheres of opposite polarity.
The angular size of the HCS amplitude depends, in particular, on the tilt angle $\alpha$ between magnetic and solar rotational axis.
The tilt angle is time dependent. It ranges from $\sim{10}^{\circ}$ during solar minimum (flat HCS) to $\sim{80}^{\circ}$ during maximum and reversal (wavy HCS).
Measurements of the tilt angle are provided by the  Wilcox Solar Observatory (WSO), since the 1970's to date, on a 10-day basis \cite{Hoeksema1995}.

The transport of \GCRs in the heliosphere is described by the Parker equation\,\cite{Moraal2013}:
\begin{equation}
\label{Eq::Parker}
\frac{\partial f}{\partial t}
=  \nabla\cdot [\mathbf{K}^{S}\cdot\nabla f ]
- (\vec{V}_{sw} + \vec{V}_D) \cdot\nabla f 
+ \frac{1}{3}(\nabla \cdot\vec{V}_{sw})\frac{\partial f}{\partial (ln R)}  
\end{equation}
where $f$ is the phase space density of \GCR particles,
$R=p/Z$ is their rigidity (momentum/charge ratio),
$\mathbf{K}^{S}$ is the symmetric part of the diffusion tensor,
$\vec{V}_{sw}$ is the solar wind speed, and $\vec{V}_{D}$ is the drift speed 
\begin{equation}\label{Eq::DriftSpeed}
  \vec{V}_{D}=  \frac{\beta{R}}{3}\nabla \times \frac{\vec{B}}{B^2} \,.
\end{equation}
The \GCR flux $J=J(t,R)$ is given by $J=\frac{\beta{c}}{4\pi}n$, where $\beta{c}$ is their speed and $n=4{\pi}R^{2}f$ is their number density.
In this work, we solved Eq.\,\ref{Eq::Parker} by means of the \emph{stochastic differential equation} method
in steady-state conditions ($\partial/\partial{t}=0$) \cite{Strauss2017},

The diffusion of \GCR particles arises from their scattering off the small-scale irregularities of the turbulent IMF.
Drift motion is caused by gradient and curvature of the regular component of the IMF, and in particular across HCS.
Diffusion and drift can be formally incorporated in the diffusion tensor $\mathbf{K}$ as symmetric and antisymmetric parts, respectively:
$\mathbf{K}=\mathbf{K}^S+\mathbf{K}^A$, with $K_{ij}^S = K_{ji}^S$ and $K_{ij}^A = -K_{ji}^A$.
However, in Eq.\,\ref{Eq::Parker}, drift is explicitly accounted by the $\vec{V}_{D}$-term, and thus only the symmetric
part of the diffusion tensor appears in the $\mathbf{K}$-term \cite{Moraal2013}.
The $\mathbf{K}^{S}$ tensor can be also split into parallel and perpendicular diffusion $K_{\parallel}$ and $K_{\perp}$,
where we assume $K_{\perp}= \xi K_{\parallel}$, with  $\xi \cong\,0.02$ \cite{Giacalone1999}.
The corresponding \emph{mean free paths} are $\lambda_{\parallel}$ and $\lambda_{\perp}$, respectively, such that
$K_{\parallel} = \beta c \lambda_{\parallel}/3$,  where $\beta=v/c$ is the particle speed.
A large compilation of observational on the parallel mean free path in the $\sim$\,0.5\,MV - 5\,GV rigidity range was reported in \citet{Palmer1982}.
The mean free path, however, is rigidity and time dependent. From the condition of cyclotron resonance,
the scattering of \GCRs occurs when their Larmor radius $r_{L}=r_{L}(R)$  is comparable with the typical size of the irregularities $\hat{\lambda}$.
From the condition $r_{L} \sim \hat{\lambda}$, it turns out that \GCRs with rigidity $R$ resonate at wave number $k_{\rm{res}} \sim 1/R$.
The IMF irregularities follows a distribution of the type $w(k) \propto k^{-\eta}$, which is 
the spectrum of interplanetary turbulence expressed in terms of wave numbers $k=2\pi/\lambda$.
An important parameter is the index $\eta$, on which different regimes can be distinguished for the IMF power spectrum \cite{Kiyani2015}. 
The resulting rigidity dependence of the diffusion mean free path (or coefficient) is $\lambda_{\parallel} \sim R^{2-\eta}$.
To account for different regimes in the IMF power spectrum  \cite{Kiyani2015},
the mean free paths are often parameterized as a \emph{double power-law} function of the particle rigidity.
For the parallel component, we have adopted the following description: 
\begin{equation}\label{Eq::Par_diff}
\lambda_{\parallel} =K_{0} \left(\frac{B_0}{B}\right) \left(\frac{R_0}{R}\right)^{a} \times\left[ \frac{(R/R_0)^h + (R_k/R_0)^h }{1 + (R_k/R_0)^h} \right]^{\frac{b-a}{h}}\,, 
\end{equation}
where $R_{0}\equiv$\,1\,GV sets the rigidity scale,  $B_{0}$ is the local value of the IMF $B$ at $r_{0}=$\,1 AU,
and the normalization factor $K_{0}$ is given in units of $10^{23}$ $cm^{2}s^{-1}$.
The spectral indices $a$ and $b$ set the slopes of the rigidity dependence of $\lambda_{\parallel}$ below and above $R_{k}$, respectively.
The parameter $h$ sets the smoothness of the transition.
The perpendicular component follows from $\lambda_{\perp}\equiv \xi \lambda_{\parallel}$,
with the addition of small corrections in the polar regions \cite{Heber1998}.

\subsection{The key parameters} 
\label{Sec::Parameters}         
%
In general, all parameters entering Eq.\,\ref{Eq::Par_diff} might be time-dependent \cite{Manuel2014}. 
We identify a minimal set of \emph{diffusion parameters} as ${K_{0}, a, b}$.
These parameters and their temporal dependence will be determined using time-resolved 
\GCR proton data from AMS-02 and PAMELA \cite{Aguilar2018PHeVSTime,Adriani2013,Martucci2018}. 
Along with diffusion parameters, we define a minimal set of \emph{heliospheric parameters} $\{\alpha, B_{0}, A\}$
that describe the time-dependent conditions of the heliosphere in a given epoch: the HCS tilt angle $\alpha$, the local IMF intensity $B_{0}$,
and its polarity $A$. Magnetic polarity is defined as the sign of the IMF
in the outward (inward) direction from the Sun's North (South) pole.
To obtain the solution of Eq.\,\ref{Eq::Parker} for a given \GCR species, the LIS has to be specified as boundary condition.
For \GCR protons, our LIS model is obtained by Galactic propagation calculations and \GCR flux
data \cite{Tomassetti2015TwoHalo,Tomassetti2012Hardening,Feng2016,Tomassetti2018PHeVSTime}.
%
\begin{figure}[hbt!]
\centering
\includegraphics[width=0.45\textwidth,scale=0.50]{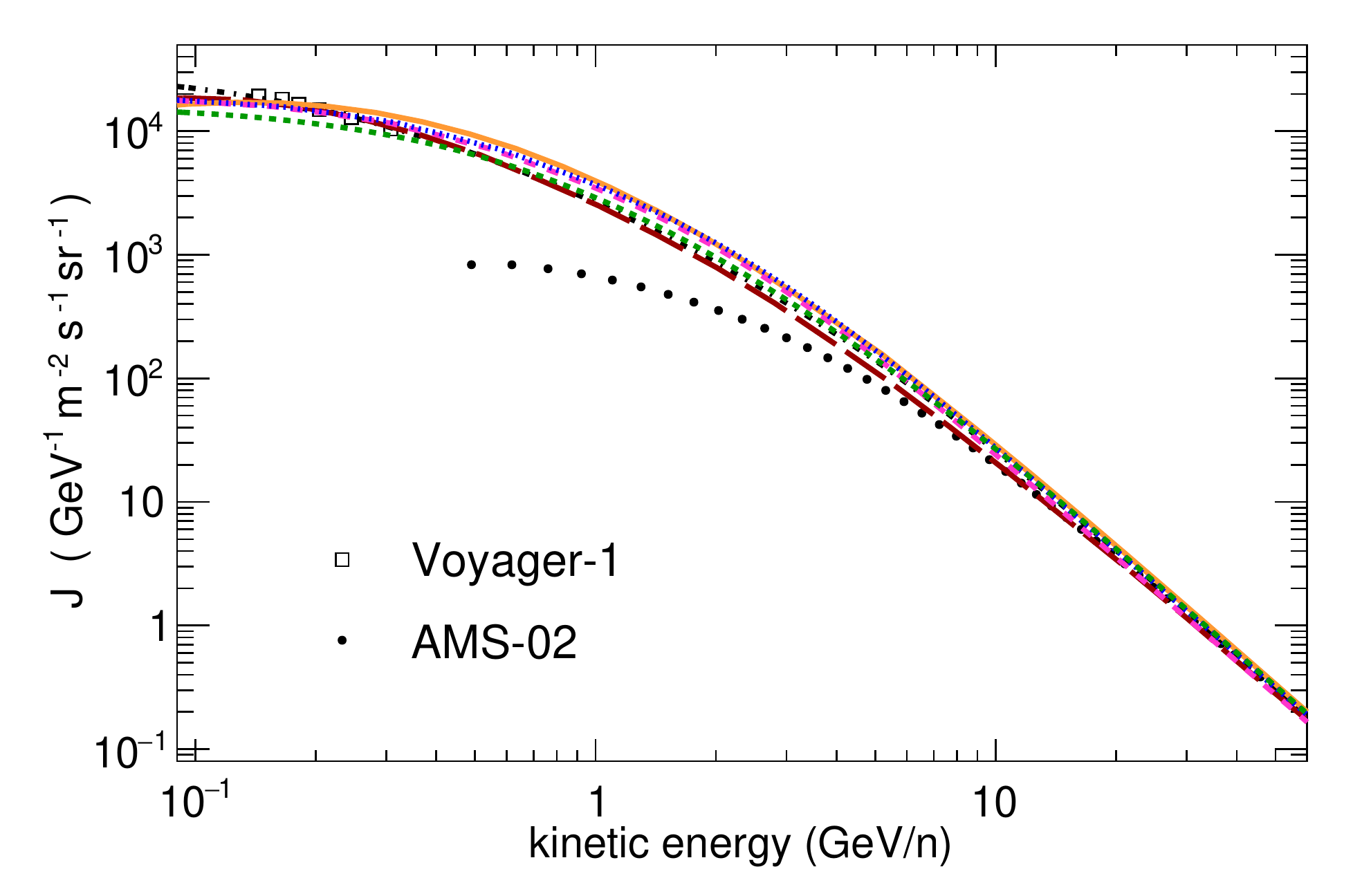} 
\caption{%
  Compilation of proton LIS modesl from various works:
  long-dashed red \citep{Tomassetti2017TimeLag},
  dot-dashed black \citep{Corti2019} 
  dotted blue \citep{Tomassetti2017BCUnc},
  dotted green \citep{Boschini2017}, 
  dashed pink \citep{Tomassetti2018PHeVSTime},
  solid orange line \citep{Tomassetti2017Universality}.
  Data are from Voyager-1 \cite{Cummings2016} and AMS-02 \citep{Aguilar2015Proton}.
}
  \label{Fig::ProtonLISModels}
\end{figure}
%
The data, used to constrain the \GCR propagation model, are from the Voyager-1 spacecraft at $\sim$\,100\,-500\,MeV of kinetic energy \cite{Cummings2016},
and from AMS-02 experiment at $E\sim$\,100\,GeV -- 2\,TeV \cite{Aguilar2018PHeVSTime,Aguilar2015Proton,Aguilar2015Helium}.
Our proton LIS agrees fairly well with other recent models
\cite{Boschini2017,Corti2019,Tomassetti2017TimeLag,Tomassetti2015PHeAnomaly,Tomassetti2017Universality,Tomassetti2017BCUnc}.
A compilation of proton LIS models is shown in Fig.\,\ref{Fig::ProtonLISModels}, along with the data from Voyager-1 and AMS-02.

\subsection{The analysis}  
\label{Sec::Analysis}      
%
We determine the time-dependent \GCR diffusion parameters by means of a statistical inference
on the monthly measurements of the \GCR proton fluxes reported by AMS-02 and PAMELA \cite{Aguilar2018PHeVSTime,Adriani2013,Martucci2018}.
For every month, however, the heliospheric parameters have to be specified as well. They are evaluated from
observations of the WSO observatory ($\alpha$, $A$) and by \emph{in situ} measurements of the ACE space probe ($B_{0}$). 
For a given epochs $t$, a  backward moving average is calculated within a time window $[t-\Delta{T}, t]$, with  $\Delta{T}=6-12$\,months.
This ensures that the average values $\hat{\alpha}$, $\hat{A}$, and $\hat{B}_{0}$ reflect the average IMF conditions sampled by \GCRs arriving Earth
at the epoch $t$ \cite{Fiandrini2021,Tomassetti2017TimeLag}. 
Hence, for each epoch, the diffusion parameters $K_{0}$, $a$, and $b$ can be determined with a global fit on the \GCR proton measurements from AMS-02 and PAMELA.
In practice, to make the fit, we have built a 6D parameter grid where each node corresponds to a 
parameter configuration $\vec{q}=$ ($\alpha$, $B_0$, $A$, $K_0$,  $a$, $b$). The grid has 938,400 nodes.
With the stochastic method, the \GCR proton spectrum $J_{m}(E, \vec{q})$ was calculated for each node of the grid at several
values of kinetic energies between 20 MeV and 200 GeV. 
%
%
\begin{figure*}[hbt!]
\centering
\includegraphics[width=0.92\textwidth,scale=0.50]{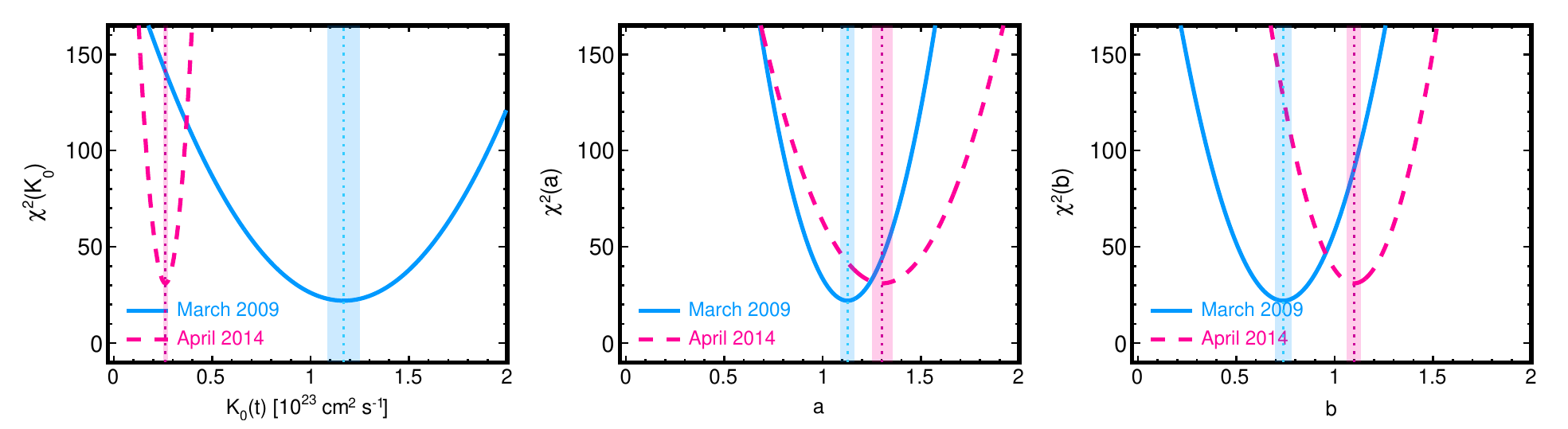} 
\caption{%
  One-dimensional projections of the $\chi^{2}$ surfaces as function of the transport parameters $K_{0}$, $a$, and $b$ evaluated
  with CR proton flux data in two epochs: April 2014, corresponding to solar maximum (pink dashed lines, from AMS-02),
  and March 2009 corresponding to solar minimum (blue solid lines, from PAMELA).
}
\label{Fig::Chi2Surface}       
\end{figure*}
%
%
The simulation was highly CPU consuming. It required the simulation of 14 billion pseudo-particle trajectories,
backward-propagated  from Earth to the heliopause and then re-weighted according to their LIS.
Once the proton grid was fully sampled, the parameters were determined as follows. 
From measured fluxes $J_{d}(E,t)$ made at epoch $t$, the model calculation $J(E,\vec{q})$ was evaluated as function of its parameters.
The heliospheric parameters ${\hat{\alpha},\hat{B_{0}},\hat{A}}$ were kept fixed at their evaluation at epoch $t$.
For a \GCR flux measurements $J_{m}(E,t)$, as function of energy $E$ and observed at epoch $t$,
the diffusion parameters are determined by the minimization of the function:
\begin{equation} \label{Eq::ChiSquare}
  \chi^{2}(K_{0},a,b) = \sum_{i}  \frac{\left[ J_{d}(E_{i},t) - J_{m}(E_{i}, \vec{q}) \right]^{2}}{\sigma^{2}(E_{i},t)} \,,
\end{equation} 
where the errors are given by $\sigma^{2}(E_{i},t) = \sigma_{d}^{2}(E_{i},t) + \sigma_{mod}^{2}(E_{i},t)$.
The errors account for various contributions:
experimental uncertainties in the data, theoretical uncertainties
of the model, and uncertainties associated with the minimization procedure.
The projections of the $\chi^{2}$ surfaces calculated for two flux measurements
are illustrated in Fig.\,\ref{Fig::Chi2Surface} as function of the \GCR diffusion parameters  $K_{0}$,  $a$, and $b$.
The figure shows two distinct epoch of solar minimum (March 2009) and  solar maximum (April 2014).
The best-fit parameter is shown in each curve together with its uncertainty band.
The data come from PAMELA (March 2009) and AMS-02 experiment (April 2014).
It can be seen that the parameters $K_{0}$ and $b$ are tightly constrained by the AMS-02 data.
The parameter $a$ is sensitive to low-rigidity data and thus it is better constrained by PAMELA.
In general AMS-02 gives larger $\chi^{2}$-values in comparison with PAMELA, but the convergence of the fit is overall good.

\section{Results and Discussion}    
\label{Sec::Results}                
%
\begin{figure}[hbt!]
\centering
\includegraphics[width=0.45\textwidth,scale=0.50]{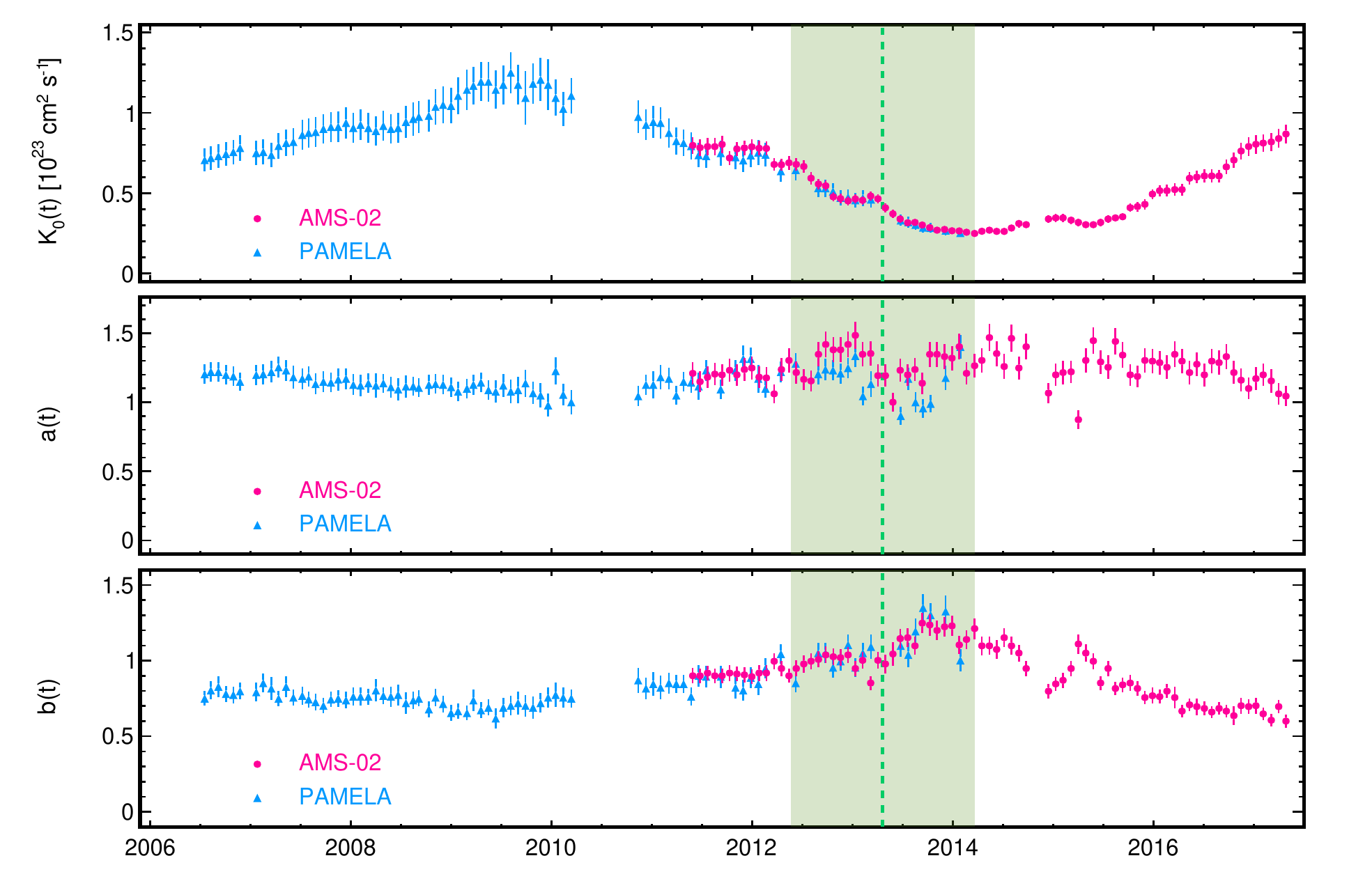}
\caption{%
  Best-fit results for the diffusion parameters ${K_{0}, a, b}$ 
  obtained with the monthly flux measurements of CR protons made by PAMELA (blue triangles) and AMS-02 (pink circles).
  The greenish band indicates the magnetic reversal epoch. During this period, the IMF polarity is not well defined.}
  \label{Fig::ccTransportParameters}
\end{figure}
%
%
\begin{figure*}[ht!] 
\centering
\includegraphics[width=0.92\textwidth,scale=0.45]{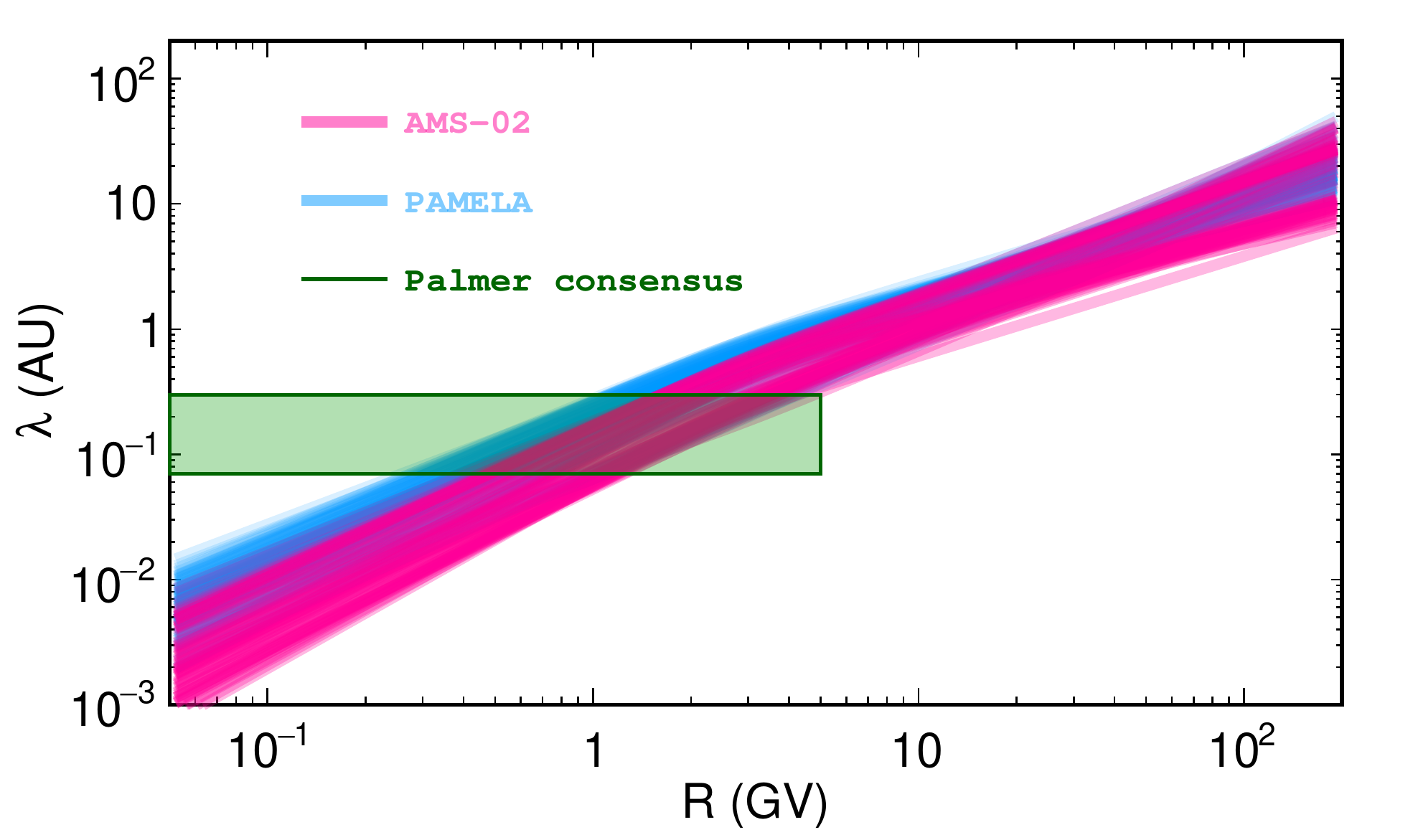}
\caption{\footnotesize{%
    Envelope of the diffusion mean free paths $\lambda_{\parallel}$ as function of \GCR rigidity inferred in the examined period.
    The pink and blue bands correspond to the AMS-02 and PAMELA dataset, respectively. Note that the two bands are largely overlapped. 
    The shaded green box corresponds to the so-called \emph{Palmer consensus} on the mean value of $\lambda_{\parallel}$ at rigidity below 5\,GV\citep{Palmer1982}.}}
\label{Fig::ccMeanFreePath}
\end{figure*}
%

Along with the two epochs of Fig.\,\ref{Fig::Chi2Surface}, the fits have been performed for the whole time-series of CR proton flux
measurements of AMS-02 and PAMELA. The AMS-02 time series consists in 79 proton fluxes measured on 27-day basis between May 2011 and May 2017.
The PAMELA series are 83 proton fluxes measured on 27-day basis between June 2006 and January 2014.
Their data, provided on monthly basis, cover large fractions of the solar cycles 23 and 24.
With the least-square minimization described in Sect.\,\ref{Sec::Analysis}, we obtained time-series
of best-fit diffusion parameters ${K_{0}, a, b}$,
along with their uncertainties. These results are shown in Fig.\,\ref{Fig::ccTransportParameters}.
The fit covers epochs of solar activity that include minimum, maximum, and IMF reversal.
From the fit, we found that the diffusion parameters show a distinct time dependence that associated with solar activity \citep{Fiandrini2021}.
The parameter $K_{0}$ is in anti-correlation with the monthly sunspot number, 
which can be understood easily within the force field model, as the modulation parameter is $\phi\propto 1/K_{0}$ \citep{Tomassetti2017BCUnc}.
Smaller $K_{0}$ values imply slower diffusion and a more significant modulation effect, \ie, a stronger suppression of the low-energy \GCR flux.
In contrast, larger $K_{0}$ values imply faster \GCR diffusion, therefore causing a minor modification of the LIS.
We found, for instance, a positive correlation between $K_{0}(t)$ and the flux $J(E_{0},t)$ evaluated at a given reference energy \citep{Fiandrini2021}.
Our finding are in good agreement with other works \citep{Manuel2014,Tomassetti2017TimeLag,Corti2019}. 
In all these works, the \GCR transport is dominated by parallel diffusion. 
We also find that the parameter $b$ shows a remarkable time dependence, reflecting the connection
between solar variability and the spectrum of magnetic turbulence in the inertial range.
In this range, the associated spectral index evolves from $\eta= 0.74\pm0.08$ at solar minimum to $\approx{1.3}\pm0.15$ during the solar maximum.
This shows that the IMF turbulence is subjected to variations  \cite{Usoskin2019,Horbury2005}.
Regarding the parameter $a$, we found a milder temporal dependence, \ie, a nearly constant spectral index of $\eta=0.79\pm0.13$.
In both ranges, our findings are in agreement with direct measurements of the IMF power spectrum \cite{Kiyani2015}.

Once the full time-series of the diffusion parameters ${K_{0}, a, b}$ is reconstructed, from their best-fit values of Fig.\,\ref{Fig::ccTransportParameters}
it is possible to calculate the time- and rigidity-dependent diffusion mean free path $\lambda_{\parallel}(t,R)$ using Eq.\,\ref{Eq::Par_diff}.
The result is shown in Fig.\,\ref{Fig::ccMeanFreePath}, where we plot the envelope of all mean free paths as function of the \GCR
rigidity inferred in the examined periods from AMS-02 (pink circles) and PAMELA (blue triangles). The two bands are largely overlapped.
The resulting mean free path for parallel diffusion in good accordance with the so-called Palmer consensus
on the observations of $\lambda_\parallel$, shown in the figure as green shaded box \citep{Palmer1982}.

\section{Acknowledgements}  
%
The present work has been developed in the framework of the joint research program between University of Perugia
and Italian Space Agency (ASI) under agreement ASI-UniPG\,2019-2-HH.0.
B. K., M. G., and F. D. acknowledges support from agreement ASI-INFN 2019-19-HH.0.
It is also acknowledged the support of Fondo Ricerca di Base of the University of Perugia.
The \GCR data used in this work have been retrieved through the cosmic-ray database of the ASI Space Science Data Center. 

\section{Declarations}  
%
All authors have approved this manuscript, agree to the order in which their names are listed, and declare that no conflict of interest exists.
The authors are responsible for the content and writing of this article.




\begin{thebibliography}{99} 



\bibitem{Moraal2013}
  Moraal, H.,
  \emph{Cosmic-Ray Modulation Equations},
  \href{https://dx.doi.org/10.1007/s11214-011-9819-3}{Space Sci Rev. 176 299 (2013)}

\bibitem{Potgieter2013}
  Potgieter, M. S.,
  \emph{Solar modulation of cosmic rays},
  \href{http://dx.doi.org/10.12942/lrsp-2013-3}{Living Rev. Solar Phys., 10, 3 (2013)}

\bibitem{Cummings2016} 
  Cummings, A. C., \etal,
  \emph{Galactic cosmic rays in the local interstellar medium: Voyager 1 observations and model results},
  \href{http://dx.doi.org/10.3847/0004-637X/831/1/18}{\ApJ{} 831, 18 (2016)} 

\bibitem{Aguilar2018PHeVSTime}
  Aguilar, M., \etal, 
  \emph{Observation of Fine Time Structures in the Cosmic Proton and Helium Fluxes with the Alpha Magnetic Spectrometer on the International Space Station},
  \href{https://journals.aps.org/prl/abstract/10.1103/PhysRevLett.121.051101}{\PRL{} 121, 051101 (2018)}

\bibitem{Aguilar2018LeptonVSTime}
  Aguilar, M., \etal, 
  \emph{Observation of Complex Time Structures in the Cosmic-Ray Electron and Positron Fluxes with the Alpha Magnetic Spectrometer on the International Space Station},
  \href{https://journals.aps.org/prl/abstract/10.1103/PhysRevLett.121.051102}{\PRL{} 121, 0511012 (2018)}

\bibitem{Adriani2013}
  Adriani, O., \etal,
  \emph{Time dependence of the proton flux measured by PAMELA during the 2006 July - 2009 December solar minimum},
  \href{https://iopscience.iop.org/article/10.1088/0004-637X/765/2/91/pdf}{\ApJ{} 765, 91, (2013)}
  
\bibitem{Martucci2018}
  Martucci, M., \etal, 
  \emph{Proton Fluxes Measured by the PAMELA Experiment from the Minimum to the Maximum Solar Activity for Solar Cycle 24}
  \href{http://dx.doi.org/10.3847/2041-8213/aaa9b2}{\ApJ{} 854, L2 (2018)}

\bibitem{Strauss2017}
  Strauss, R. D., Effenberger, F.,
  \emph{A Hitch-hiker’s Guide to Stochastic Differential Equations. Solution Methods for Energetic Particle Transport in Space Physics and Astrophysics},
  \href{http://dx.doi.org/10.1007/s11214-017-0351-y}{Space Sci Rev (2017)}

\bibitem{Kappl2016}
  Kappl, R.,
  \emph{SOLARPROP: Charge-sign dependent solar modulation for everyone},
  \href{https://doi.org/10.1016/j.cpc.2016.05.025}{Comp. Phys. Comm. 207 (2016) 386-399}

\bibitem{Tomassetti2017BCUnc}
  Tomassetti, N.,
  \emph{Solar and nuclear physics uncertainties in cosmic-ray propagation},
  \href{http://dx.doi.org/10.1103/PhysRevD.96.103005}{\PRD{} 96, 103005 (2017)}

\bibitem{Fiandrini2021}
  Fiandrini, E., Tomassetti, N., Bertucci, \etal,
  \emph{Numerical modeling of cosmic rays in the heliosphere: Analysis of proton data from AMS-02 and PAMELA},
  to appear in \PRD{} \href{}{arXiv:2010.08649} (2021)

\bibitem{Hoeksema1995}
  Hoeksema, J. T.,
  \emph{The large-scale structure of the heliospheric current sheet during the Ulysses epoch},
  \href{https://doi.org/10.1007/BF00768770}{Space Sci. Rev. 72, 137-148 (1995)}

\bibitem{Giacalone1999}
  Giacalone, J., \& Jokipii, J. R., 
  \emph{The Transport of Cosmic Rays across a Turbulent Magnetic Field},
  \href{https://doi.org/10.1086/307452}{\ApJ{} 520, 204 (1999)}

\bibitem{Kiyani2015}
  Kiyani K. H., Osman K. T., Chapman S.C., 
  \emph{Dissipation and heating in solar wind turbulence: from the macro to the micro and back again},
  \href{http://dx.doi.org/10.1098/rsta.2014.0155}{Phil. Trans. R. Soc. A 373: 20140155 (2015)} %

\bibitem{Heber1998}
  Heber, B., \etal, 
  \emph{Latitudinal distribution of greater than 106 MeV protons and its relation to the ambient solar wind in
  the inner southern and northern heliosphere - ULYSSES Cosmic and Solar Particle Investigation Kiel Electron Telescope results},
  \href{http://dx.doi.org/10.1029/97JA01984}{J. Geophys. Res., 103, 4809 (1998)}

\bibitem{Manuel2014}
  Manuel, R., Ferreira, S. E. S., Potgieter, M. S.,
  \emph{Time-Dependent Modulation of Cosmic Rays in the Heliosphere},
  \href{http://dx.doi.org/10.1007/s11207-013-0445-y}{Sol. Phys. 289, 2207 (2014)} 

\bibitem{Tomassetti2015TwoHalo}
  Tomassetti, N.,
  \emph{Cosmic-ray protons, nuclei, electrons, and antiparticles under a two-halo scenario of diffusive propagation},
  \href{http://dx.doi.org/10.1103/PhysRevD.92.081301}{\PRD{} 92, 081301 (2015)} 

\bibitem{Tomassetti2012Hardening}
  Tomassetti, N.
  \emph{Origin of the cosmic ray spectral hardening},
  \href{http://dx.doi.org//10.1088/2041-8205/752/1/L13}{\ApJ{} Lett. 752, L13 (2012)} 
  
\bibitem{Feng2016}
  Feng, J., Tomassetti, N., Oliva, A.,
  \emph{Bayesian analysis of spatial-dependent cosmic-ray propagation: Astrophysical background of antiprotons and positrons},
  \href{http://dx.doi.org/10.1103/PhysRevD.94.123007}{\PRD{} 94, 123007 (2016)}
  
\bibitem{Tomassetti2018PHeVSTime}
  Tomassetti, N., Bertucci, B., Bar\~{a}o, F., \etal,
  \emph{Testing Diffusion of Cosmic Rays in the Heliosphere with Proton and Helium Data from AMS},
  \href{http://dx.doi.org/10.1103/PhysRevLett.121.251104}{\Journal{\PRL}{121}{251104}{2018}}

\bibitem{Aguilar2015Proton} 
  Aguilar, M., \etal, 
  \emph{Precision Measurement of the Proton Flux in Primary Cosmic Rays from Rigidity 1 GV
    to 1.8 TV with the Alpha Magnetic Spectrometer on the International Space Station},
  \href{http://dx.doi.org/10.1103/PhysRevLett.114.171103}{\Journal{\PRL}{114}{171103}{2015}}

\bibitem{Aguilar2015Helium}
  Aguilar, M., \etal, 
  \emph{Precision Measurement of the Helium Flux in Primary Cosmic Rays of Rigidities 1.9 GV
  to 3 TV with the Alpha Magnetic Spectrometer on the International Space Station},
  \href{http://dx.doi.org/10.1103/PhysRevLett.115.211101}{\Journal{\PRL}{115}{211101}{2015}}

\bibitem{Boschini2017}
  Boschini, M. J., \etal, 
  \emph{Solution of Heliospheric Propagation: Unveiling the Local Interstellar Spectra of Cosmic-ray Species},
  \href{http://dx.doi.org/10.3847/1538-4357/aa6e4f}{\ApJ{} 840, 115 (2017)} 
  
  
\bibitem{Corti2019}
  Corti, C., \etal, 
  \emph{Numerical modeling of galactic cosmic ray proton and helium observed by AMS-02 during the solar maximum of Solar Cycle 24},
  \href{http://dx.doi.org/10.3847/1538-4357/aafac4}{\Journal{\ApJ}{871}{253}{2019}}
  

\bibitem{Tomassetti2017TimeLag}
  Tomassetti, N., Orcinha, M., Bar\~{a}o, F., Bertucci, B., 
  \emph{Evidence for a Time Lag in Solar Modulation of Galactic Cosmic Rays},
  \href{http://dx.doi.org/10.3847/2041-8213/aa9373}{\ApJ{} Lett. 849, 32 (2017)} 
  

\bibitem{Tomassetti2015PHeAnomaly}
  Tomassetti, N.,
  \emph{Origin of the proton-to-helium ratio anomaly in cosmic rays}, 
  \href{http://dx.doi.org/10.1088/2041-8205/815/1/L1}{\ApJ{} Lett. 815, L1 (2015)} 
  
\bibitem{Tomassetti2017Universality}
  Tomassetti, N.,
  \emph{Testing universality of cosmic-ray acceleration with proton/helium data from AMS and Voyager-1}
  \href{http://dx.doi.org/10.1016/j.asr.2016.10.024}{\ASR{} 60, 815-825 (2017)} 
  
\bibitem[Palmer(1982)]{Palmer1982}
  Palmer, I. D.,
  \emph{Transport coefficients of low-energy cosmic rays in interplanetary space},
  \href{https://doi.org/10.1029/RG020i002p00335}{Rev. Geophys. Space Phys. 20, 335 (1982)} 

\bibitem{Usoskin2019} 
  Vaisanen, P., Usoskin, I., Mursula, K.,
  \emph{Long-Term and Solar Cycle Variation of Galactic Cosmic Rays: Evidence for Variable Heliospheric Turbulence},
  \href{https://doi.org/10.1029/2018JA026135}{J. Geophys. Res.: Space Phys. 124, 804-811 (2019)} 
  
\bibitem{Horbury2005}
  Horbury, T. S., Forman, M. A., \& Oughton, S.,
  \emph{Spacecraft observations of solar wind turbulence: an overview},
  \href{https://doi.org/10.1088/0741-3335/47/12B/S52}{Plasma Phys. Contr. Fusion, 47, B703-B717 (2005)} 

  
 

\end{thebibliography}
\end{document}